\documentclass[12pt]{article}

\usepackage{scicite}
\usepackage{times}
\usepackage{graphicx,amsfonts,amssymb,amsmath}

	\title{Self-supervised Learning and Prediction of Microstructure Evolution with Recurrent Neural Networks}

\author
{Kaiqi Yang,$^{1}$ Yifan Cao,$^{1}$ Youtian Zhang,$^{1}$ Ming Tang,$^{1\ast}$ \\ Daniel Aberg,$^{2}$ Babak Sadigh,$^{2}$  Fei Zhou$^{2\ast}$\\
	\\
	\normalsize{$^{1}$Department of Materials Science and NanoEngineering, Rice University, Houston, TX 77005, USA}
	\\
	\normalsize{$^{2}$Physical and Life Sciences Directorate, Lawrence Livermore National Laboratory,}\\ \normalsize{Livermore, CA 94550, USA}\\
	\\
	\normalsize{$^\ast$To whom correspondence should be addressed; E-mail: zhou6@llnl.gov, mt20@rice.edu.}
}

\date{}

\begin{document} 
	
	
	\baselineskip24pt
	
	
	\maketitle

	

%
%
%
%

	\begin{abstract}	
Microstructural evolution is a key aspect of understanding and exploiting the structure-property-performance relation of materials. Modeling microstructure evolution usually relies on coarse-grained simulations with evolution principles described by partial differential equations (PDEs). Here we demonstrate that convolutional recurrent neural networks can learn the underlying physical rules and replace PDE-based simulations in the prediction of microstructure phenomena. Neural nets are trained by self-supervised learning with image sequences from simulations of several common processes, including plane wave propagation, grain growth, spinodal decomposition and dendritic crystal growth. The trained networks can accurately predict both short-term local dynamics and long-term statistical properties of microstructures and is capable of extrapolating beyond the training datasets in spatiotemporal domains and configurational and parametric spaces. Such a data-driven approach offers significant advantages over PDE-based simulations in time stepping efficiency and offers a useful alternative especially when the material parameters or governing PDEs are not well determined.  
	\end{abstract}

\section*{Introduction}
Materials microstructures are mesoscale structural features that serve as an indispensable link between the atomistic building blocks and macroscopic properties, leading to direct impact on the processing-structure-property relationship of engineered materials.
Tailoring materials properties through controlled microstructure evolution under non-equilibrium conditions during materials processing or service,
 including ubiquitous phenomena such as solidification, solid-state phase transformations and grain growth, is arguably a cornerstone of modern materials science. The ability to understand and predict microstructure evolution has therefore long been a pivotal goal of computational materials design. 

Due to time and length scales well beyond the capability of molecular dynamics,
simulations of microstructure evolution often rely on coarse-grained models such as partial differential equations (PDEs) as employed in the phase-field method \cite{Phase-Field-Provatas}.  Nevertheless, this approach faces several significant difficulties. First, while in principle PDEs can be derived from the underlying thermodynamic and kinetic considerations,  identifying, parametrizing and validating PDEs in practice require intricate evolution rules and substantial manual effort. For complicated or less studied materials, the evolution rules might be either not fully understood or too complex to be described by tractable PDEs. Second, microstructure simulations employing PDEs remain fairly expensive. In the temporal dimension, strict upper limits on the minimum time step size are dictated by the stability of numerical schemes for nonlinear PDEs. In the spatial dimensions, the small grid size determined by the physical interface width often makes direct numerical simulations so expensive that special treatments such as the diffuse interface were introduced \cite{Phase-Field-Provatas}. 

We propose an alternative machine-learning (ML) method to microstructure evolution modeling. Recent progress in ML and deep neural networks\cite{LeCun2015N436} in particular enables a data-driven approach to solving PDEs in place of traditional numerical methods\cite{Raissi2017a1711.10561, Raissi2017a1711.10566, Brunton2016PNAS3932,Rudy2017SAe1602614, Champion2019PNAS22445, Schaeffer2017PRSA20160446, Breen2019a1910.07291}.  
Based on statistical learning with big datasets, ML models can be applied without explicit prior knowledge of the physical mechanisms.
With proper training, it is possible for ML algorithms to infer ``hidden'' parameters from the input microstructure images and identify the correct evolution trajectory.
Moreover, ML models allow much larger time stepping to achieve significant speedup in the temporal domain.
For example, Raissi and coworkers used a single four-layer neural netwrok\cite{Raissi2017a1711.10561, Raissi2017a1711.10566} to obtain the solutions to the Burger's equation, which otherwise require 500 Runge-Kutta iterations. 
Breen {\it et al.} tackled the notoriously difficult three-body problem with a ten-layer neural nets, skipping thousands of smaller time steps\cite{Breen2019a1910.07291}. 
Similarly, coarser spatial grids may be used, as will be shown later.
Although previous studies reveal the power of neural nets in rediscoveringhttps://www.overleaf.com/project/5f361814eb30ee00018f96f0 and solving different types of differential equations,
they are mainly limited to ordinary differential equations  and PDEs in 1$+$1 dimensions (i.e. 1 spatial and 1 temporal dimensions). 
Deep learning of microstructure evolution, which requires PDEs in 2$+$1 or 3$+$1 dimensions, 
remains a challenging subject.

In this work, we apply the recurrent neural networks (RNN)
to predict the spatiotemporal evolution of microstructure represented by two-dimensional (2D) image sequences. 
RNNs are neural nets designed to predict temporal data sequence with hidden memory units\cite{10.5555/65669.104451,werbos1990backpropagation}.
With the development of effective variants such as the long short term memory (LSTM) to address the vanishing gradient problem during backpropagation\cite{hochreiter1997long},
RNNs have found wide-spread success in natural language processing \cite{sutskever2011generating,cho2014properties}, 
speech recognition
\cite{graves2014towards} 
and computer vision\cite{srivastava2015unsupervised,donahue2015long,finn2016unsupervised}.
In recent years, several variants of LSTM combined with convolutional neural nets (CNN) have been proposed for predictive learning of spatiotemporal sequences, including the convolutional LSTM (ConvLSTM)\cite{xingjian2015convolutional}, Predictive RNN (PredRNN)\cite{Wang2017PredRNN}, PredRNN++\cite{Wang2018PredRNN++} and eidetic 3D LSTM (E3D-LSTM)\cite{Wang2019E3D-LSTM}. 
These models employ CNN for efficient spatial latent-feature extraction and LSTM for feature time evolution to make full use of features in both spatial and temporal domains.
We choose the more recent E3D-LSTM method for this study and use the terms E3D-LSTM and RNN interchangeably hereafter.

We assess RNN's learning ability and predictive power in the context of four well-known evolution phenomena with increasing level of complexity: plane wave propagation, grain growth, spinodal decomposition and dendritic crystal growth.
To facilitate comparison with physics-based models, the training datasets are generated from PDE-based simulations or explicit mathematical functions, whose behavior is well understood. 
A focus of our study is to examine to what degree RNN can grasp and extract the evolution rules from the microstructure images it sees.
To this end, extensive and stringent tests are devised to evaluate how well RNN generalizes and extrapolates the learning within the spatiotemporal domain and configurational and parametric spaces.
We find that properly trained RNN is able to extend the predictions up to ten folds of the time spans of the training data with significantly larger time step sizes than used in training PDEs, and to systems of larger dimensions. 
It can forecast the evolution of systems with underlying material parameters or initial configurations that are significantly different from the training images.
In addition to excellent piece-wise comparison between the ground truth and short-term predictions,
RNN accurately captures the statistical properties of microstructures in the examples considered in the long-term.    
The satisfactory performance of RNN in these tests provides compelling evidence that it is capable of ``comprehending'' the physical principles underlying diverse microstructure evolution phenomena,
which explains why it is able to make reliable predictions well beyond the scope of training data. 
Such extrapolation capability further improves the efficiency of RNN by allowing it to be trained with relatively small data size.  
Our work illustrates the promise of ML approaches in general as a useful alternative to physics-based simulations of microstructure evolution. 

Broadly speaking, the use of ML algorithms is growing very rapidly in materials science in recent years
 \cite{schmidt2019recent, rickman2019materials,ramprasad2017machine,butler2018machine}.
 They have seen diverse applications ranging from the discovery of new materials
 \cite{ryan2018crystal,graser2018machine,balachandran2018experimental,ye2018deep,faber2016machine} to the predictions of materials properties
 \cite{xie2018crystal,isayev2017universal,yuan2017identifying,kim2016machine,carrete2014finding}, the development of accurate and efficient potentials for atomistic simulations
 \cite{pun2019physically,behler2016perspective, botu2015learning,behler2007generalized}, and microscopic and spectroscopic data analysis and processing
 \cite{decost2017exploring,decost2015computer,chowdhury2016image,ling2017building,azimi2018advanced,stan2020optimizing,bulgarevich2018pattern,rickman2017data,ziatdinov2017learning,ziatdinov2017deep,ding2019joint,mao2019high,zheng2020random}.
 A large number of these works are devoted to materials microstructure, including microstructure classification
 \cite{decost2017exploring, decost2015computer,chowdhury2016image,ling2017building,azimi2018advanced}, image segmentation
 \cite{stan2020optimizing,bulgarevich2018pattern}, 
 predictions of microstructure-property relations
 \cite{rickman2017data,gusenbauer2020extracting,kondo2017microstructure,cecen2018material} and microstructure optimization
 \cite{liu2015predictive,liu2017artificial,exl2018magnetic} with encouraging results. 
 Datasets in these works are mainly in the form of static microstructure images. This work reveals important critical temporal correlation between images of microstructures in their time evolution trajectory.

\section*{Results}
We employ numerical simulations to generate sequences of 64$\times$64-pixel images as training datasets for four classical examples of evolution phenomena, i.e. plane wave propagation, grain growth, spinodal decomposition and dendritic crystal growth. 
With varied complexity, they represent a good combination of testing problems for evaluating the capability of RNN in predicting microstructure evolution.

\subsection{Plane wave propagation}
Before delving into problems pertinent to real materials, 
we first test RNN with a simple toy model -- plane wave propagation dynamics of a scalar field $c$ explicitly described by the following expression: 
\begin{equation}\label{eq:wave}
    c(x,y,t) = \frac{1}{2} \sin(k_x x+k_y y+\omega t + \theta_0) \exp(-\beta t) + \frac{1}{2}
\end{equation}
where $\vec{k}$=($k_x$, $k_y$) is the wave vector, $\theta_0$ is a random phase and $\beta$ is a decay exponent.  
We use Eq.~\ref{eq:wave} to generate image sequences, each of which consists of 200 frames at a time interval of 0.005 between two adjacent frames starting at $t=0$. 
The parameters in Eq.~\ref{eq:wave} are randomly chosen for each sequence: $2\pi/|\vec{k}| \in [0.3,0.6]$, $2\pi/\omega \in [0.03, 0.06]$, $2\pi/\beta \in [1.5, 6]$ and $\theta_0 \in [0, 2\pi]$. 
Among the generated sequences, 80 are used for training, 20 for validation and 100 for testing. 
Each simulation sequence is divided into staggered 20-frame training clips (i.e. frame 1--20, 11--30, etc),
each of which represents a training data point. 
For testing, RNN is used to predict the next 50 frames based on an input of 10 consecutive frames. 
A total of 1500 tests are performed.  
\begin{figure}[htp]
	\centering
	\includegraphics[width=1\columnwidth]{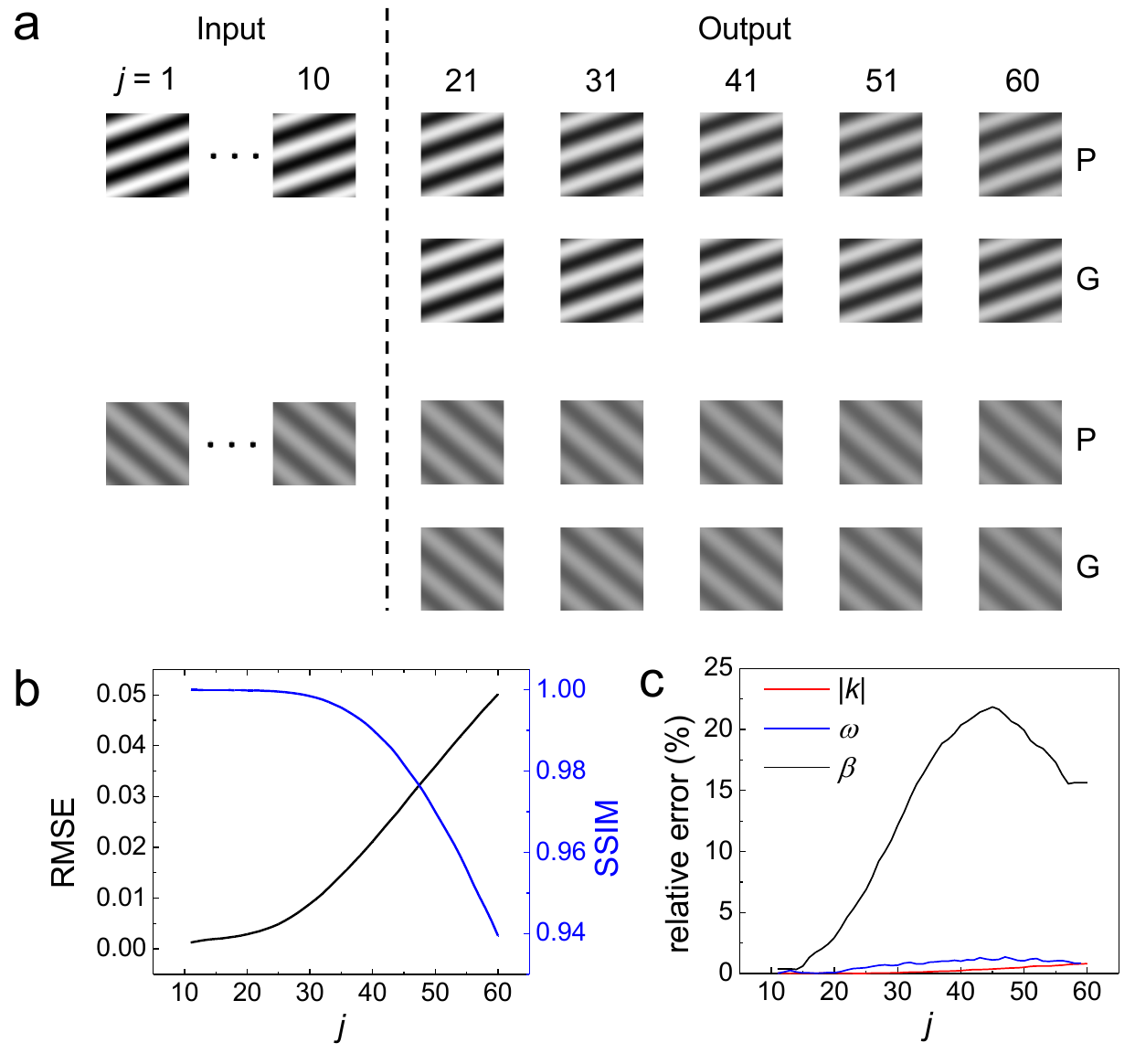}
	\caption{\textbf{Application of RNN to predicting plane wave propagation.} 
	\textbf{a.} Examples of output frames predicted by the trained RNN (P) based on 10 input frames in comparison with the ground truth (G). 
	\textbf{b.} RMSE (black) and SSIM (blue) of the predictions averaged over 200 testing cases as a function of the frame index $j$.
	\textbf{c.} Relative errors of the wave propagation parameters ($|k|$, $\omega$ and $\beta$) inferred from the predicted images. }
	\label{fig:wave1}
\end{figure}

\textbf{Figure} \ref{fig:wave1}a illustrates two representative tests, which visually show little difference between the ground truth and predictions. 
\textbf{Figure} \ref{fig:wave1}b shows the pixel-wise comparison based on the root-mean-squared-error (RMSE) and structural similarity index measure (SSIM)\cite{wang2004image} averaged over all of the 1500 tests.
Both RMSE and SSIM vary between 0 and 1 and lower RMSE or higher SSIM scores indicate better agreement between the predictions and ground truth.
It can be seen that RNN exhibits high piece-wise accuracy in the short-term within the length of training clips, where RMSE stays below 0.5\% and SSIM above 99\%. 
In the longer term, both RMSE and SSIM vary with time at a greater rate, but remains below 5\% (or above 93\%) up to 50 output frames.
As a more revealing measurement of how well RNN recognizes the wave propagation rules,  
the parameters in Eq.~\ref{eq:wave} are extracted from the predicted images and compared with their ground truth values. 
As shown in \textbf{Figure} \ref{fig:wave1}d, the predicted $|\vec{k}|$ and $\omega$ differ from the ground truth by less than 2\% but $\beta$ shows a larger deviation up to 20\%. 
A probable reason for the predicted $\beta$ being less accurate is that $\beta$ characterizes a slower decaying mode of wave motion and may require longer training sequences to learn precisely its temporal behavior.

Overall, RNN exhibits excellent performance when applied to the simple plane wave propagation problem. 
Next, we test it against more realistic microstructure evolution problems.

\subsection{Grain Growth}
Grain growth describes the increase of the average grain size in polycrystals with time to reduce the excess energy associated with grain boundaries. 
During the process, some grains grow while others shrink and disappear, leading to a persisting drop of the number of grains in the system.
The growth or shirinkage rate of a grain in 2D polycrystals is determined by its number of sides $N$ according to the famous von Neumann-Mullins or ``$N$-6'' rule\cite{von1952metal, mullins1956two}:
\begin{equation}\label{eq:NeumannMullins}
	\frac{dA}{dt} = M\gamma\frac{\pi}{3}(N-6)
\end{equation}
where $A$ is the grain area, $M$ and $\gamma$ are the grain boundary mobility and energy, respectively. 
Eq.~\ref{eq:NeumannMullins} states that any grains with fewer than 6 neighbors will shrink and those with more than 6 sides will grow at a rate proportional to $N-6$.   

We generate the training data by performing isotropic 2D grain growth simulations with a phase-field model\cite{Moelans2008PRB} (see Methods).
Simulations are performed on a 256$\times$256 grid with periodic boundary conditions to accommodate a sufficient number of grains.
Subsequently, the simulation images are down-sampled to 64$\times$64 pixels by averaging. 
Each simulation employs the same parameters but starts with a different initial configuration constructed by Voronoi tessellation with 100 random seeds.
It outputs a 20-frame clip after a relaxation period, which serves to remove the artifacts in the polycrystalline structure.
The time interval between two adjacent frames corresponds to 80 PDE time steps. 
The first frame in a clip contains $\sim$75 grains and the last one has $\sim$45 grains. 
A total of 2400 clips are prepared for training and 600 for validation during training. 
\begin{figure}[!thp]
	\vskip 0.15in
	\centering
	\includegraphics[width=0.7\columnwidth]{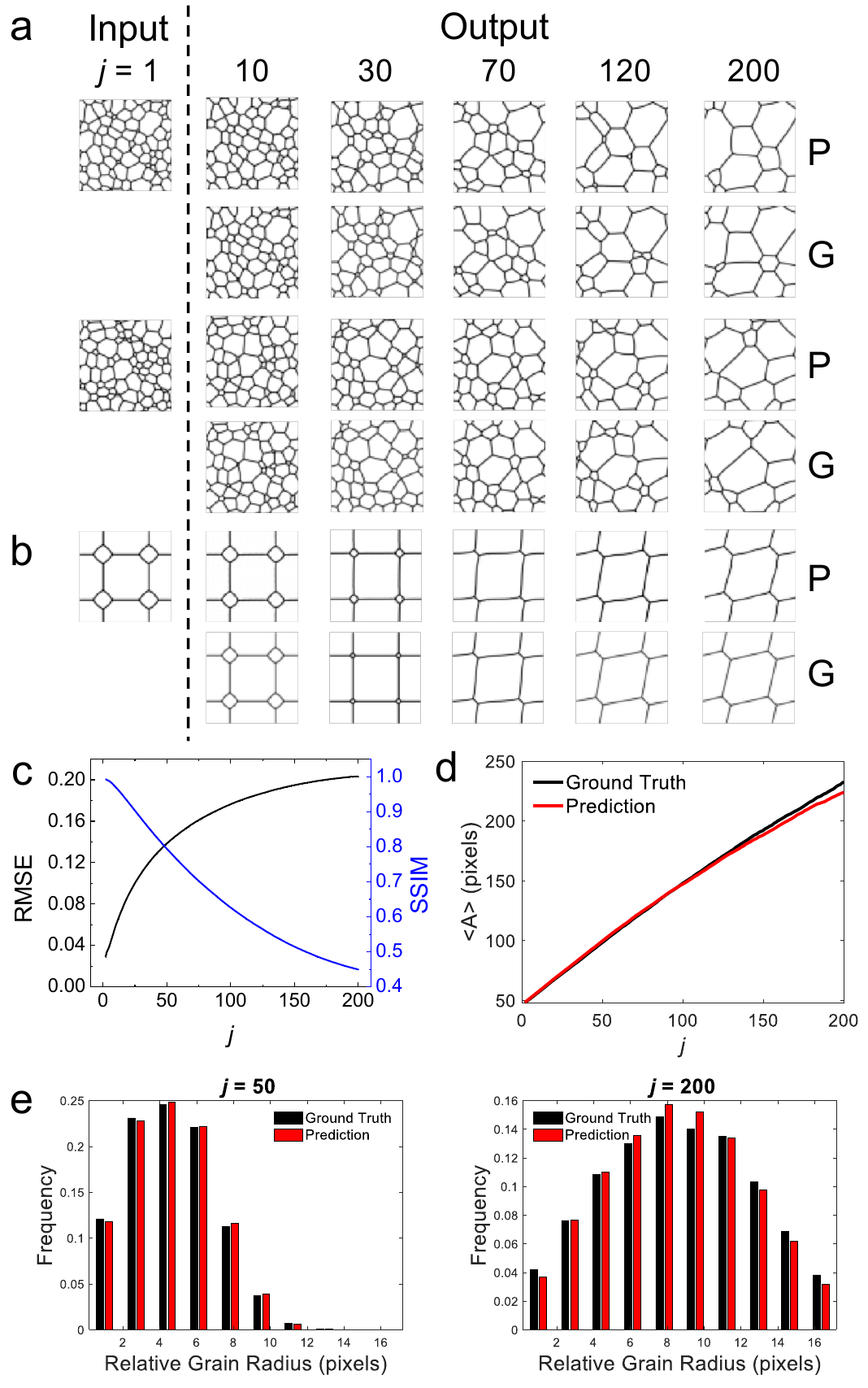}
	\caption{\textbf{Application of RNN to predicting grain growth.} 
	\textbf{a.}  Examples of RNN output frames (P) based on 1 input frame in comparison with the ground truth (G). 
	\textbf{b.} RNN prediction of the evolution of an artificial polycrystalline configuration, in which four small 4-sided grains are embedded in larger 8-sided grains.
	\textbf{c.} RMSE (black) and SSIM (blue) of the predictions averaged over 1000 cases as a function of the frame index $j$. 
	\textbf{d.} Time evolution of the average grain area in 1000 testing cases predicted by RNN vs ground truth. 
	\textbf{e.} Grain size distribution at $j$ = 50 and 200 predicted by RNN vs ground truth. Effective grain radius is calculated by $\sqrt{A/\pi}$.   
	}
	\label{fig:graingrowth1}
\end{figure}

After training, RNN is subject to a set of more challenging extrapolation tests than in the wave propagation problem. 
First, we apply the trained model to predict longer image sequences with less input information. 
RNN is required to predict 199 frames based on only one input frame.
Theoretically, this is feasible as grain growth obeys the dissipation dynamics described by PDEs of first order in time (Eq.~\ref{eq:gg3}).
Here the length of the test sequences is ten times of the training clips, and more significantly,
90\% of the output frames (frame index $j$ = 21 -- 200) depict coarsened polycrystalline states never seen by RNN during training. 
\textbf{Figure}~\ref{fig:graingrowth1}a presents two representative tests, which show that RNN does a very good job in the temporal extrapolation. 
The predictions and ground truth are difficult to distinguish visually in the short term, e.g. at frame index $j$ = 30,  
but visible local structure difference emerges at the later stage.
\textbf{Figure}~\ref{fig:graingrowth1}c shows that the average RMSE of 1000 tests rises and stabilizes around 20\% while SSIM decreases to $\sim$0.4 at the 200th frame.   
Despite the increasing difference, the predicted polycrystalline structures are free of any noticeable artifacts throughout the sequences. 
We note that the accumulation of the discrepancy between the ground truth and predictions is inevitable in the long term.
This is because the grain boundary connectivity bifurcates upon grain disappearance (see examples in Supplementary Figure S1), which leads two initially identical configurations onto divergent evolution pathways.   
As such, statistical measurement of the similarity between two polycrystalline configurations is more meaningful than pixel-wise comparison,
and RNN performs very well in this aspect. 
As shown in \textbf{Fig.}~\ref{fig:graingrowth1}d, the error in the predicted average grain area $\langle A \rangle$ of 1000 testing cases remain below 5\% while $\langle A \rangle$ has a five-fold increase.
\textbf{Figure}~\ref{fig:graingrowth1}e shows that the predictions and ground truth also have very good agreement in the grain size distribution. 
The Euclidean distance between them is only 0.71\% at $j$ = 50 and still has a low value of 1.61\% at $j$ = 200. 
RNN thus faithfully reproduces the statistical characteristics of polycrystals even after a 10-fold extrapolation in time. 

Next, we subject RNN to spatial extrapolation tests by asking it to predict grain growth in a system much larger than the training images. 
Because of the locality of 3D convolution operations in E3D-LSTM, the evolution rules learned by the model can be easily extended to larger domains without additional training. 
Supplementary Figure S2 presents the results of the grain growth kinetics on a 256$\times$256 mesh predicted by RNN trained on 64$\times$64-pixel images.
The predictions exhibit similar RMSE and SSIM compared to those for the smaller 64$\times$64-pixel domain.
The spatial extensibility of RNN means that there is no need to retrain the model when applying it to problems of different sizes, which is a very appealing feature for practical applications. 

As the third type of extrapolation tests, RNN is applied to predict the evolution of artificial polycrystalline configurations qualitatively different from the training data. 
\textbf{Figure} \ref{fig:graingrowth1}b showcases such an example, in which the system contains four orderly arranged 4-sided grains embedded within four larger grains. 
Despite the notable morphological difference from those generated by random Voronoi tessellation, the evolution of the polycrystalline is accurately captured by RNN.  
\begin{figure}[!thp]
	\vskip 0.15in
	\centering
	\includegraphics[width=0.65\columnwidth]{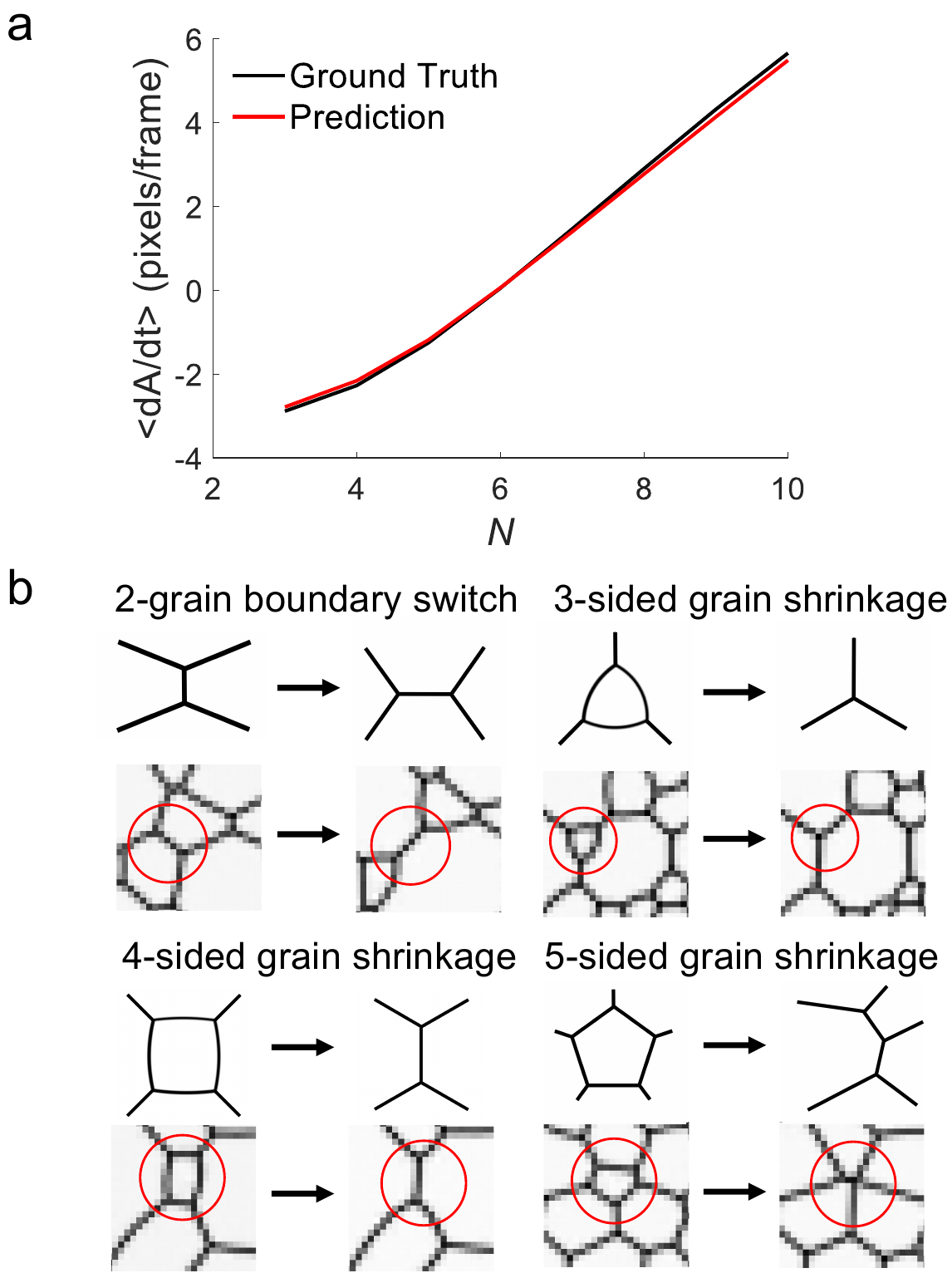}
	\caption{ \textbf{Evidence of RNN capturing the evolution rules of grain growth.}
		\textbf{a.} RNN accurately predicts the dependence of the grain growth rate $\langle dA(N)/dt\rangle$ on the number of grain sides $N$. $\langle dA(N)/dt\rangle$ is averaged over grains of the same $N$ in all of the testing cases. 
		\textbf{b.} Examples from testing cases show that RNN correctly predicts the four possible topological events when a grain disappears or loses an edge to its neighbors. Red circles highlight where the events occur in the predicted images. 
		}
	\label{fig:graingrowth2}
\end{figure}

The above tests demonstrate RNN's capability to generalize and extrapolate its learning in the spatiotemporal and configurational spaces. 
This is a strong indication that it has grasped the evolution rules, which is further supported by other evidence.
Grain growth consists of two elementary processes: the continuous shrinkage or expansion of grains without changing their number of sides $N$, 
and the discontinuous changes in the grain boundary connectivity when grains switch edges or disappear.    
The former process is governed by the $N-6$ rule (Eq.~\ref{eq:NeumannMullins}) resulting from the curvature-driven boundary movement.
In \textbf{Fig.}~\ref{fig:graingrowth2}a, we show the average growth rates for grains with different $N$ using data from all the 1000 tests.  
The predictions very faithfully reproduce the $N$ dependence of the ground truth. 
On the other hand, \textbf{Fig.}~\ref{fig:graingrowth2}b illustrates all of the four possible topological events that could occur the grain boundary network upon grain disappearance or edge switching in a 2D system. 
The numerical examples in \textbf{Fig.}~\ref{fig:graingrowth2}b show that RNN correctly predicts each one of them.
Therefore, the satisfactory performance of RNN derives from its faithful learning of the elementary steps of the grain growth process.

\subsection{Spinodal decomposition}
As a third example of microstructure evolution phenomena, we train RNN to predict spinodal decomposition, 
which is the phenomenon of spontaneous phase separation in unstable binary or multi-component systems widely found in alloys and polymer blends\cite{balluffi2005kinetics}.
Mathematically, the spatiotemporal evolution during spinodal decomposition is described by the Cahn-Hilliard (C-H) equation\cite{cahn1958free} (Eq.~\ref{eq:cahn1} in Methods),
which is numerically solved to generate the ground truth data in this work.
Compared to grain growth, spinodal decomposition is a more complex evolution phenomenon since it involves not only curvature-driven interface migration but also coupled long-range diffusion of chemical species. 
The complexity is also reflected by the 4th-order nonlinear C-H equation versus the second-order phase-field PDEs for grain growth.

Spinodal decomposition consists of two distinct stages:
a fast composition modulation growth stage, followed by a slower coarsening stage, at which the length scale of the phase separation pattern gradually increases due to the Gibbs-Thomson effect\cite{balluffi2005kinetics}.
We focus on training RNN to recognize the system evolution in the second coarsening stage. 
Training and validation data are generated from 480 and 120 simulations, respectively, 
which employ the same parameters but different initial states. 
The system starts from a uniform binary mixture with one of three compositions at $c_0$ = 0.25, 0.5 and 0.75, 
which produce different types of domain morphologies.  
A random noise of the same amplitude is added to the initial configurations to trigger phase separation.
Each simulation produces 100 images, and the system becomes phase separated after 2 or 3 frames. 
Similar to the wave propagation problem, these frames are divided into staggered 20-frame training clips (i.e. frame 1--20, 11--30, ..., and 81--100). 
The time interval between 2 adjacent frames corresponds to 370 time steps on average in phase-field simulations, 
which employ an implicit PDE solver with variable time step size.  
\begin{figure}[!thp]
	\vskip 0.15in
	\centering
	\includegraphics[width=0.8\columnwidth]{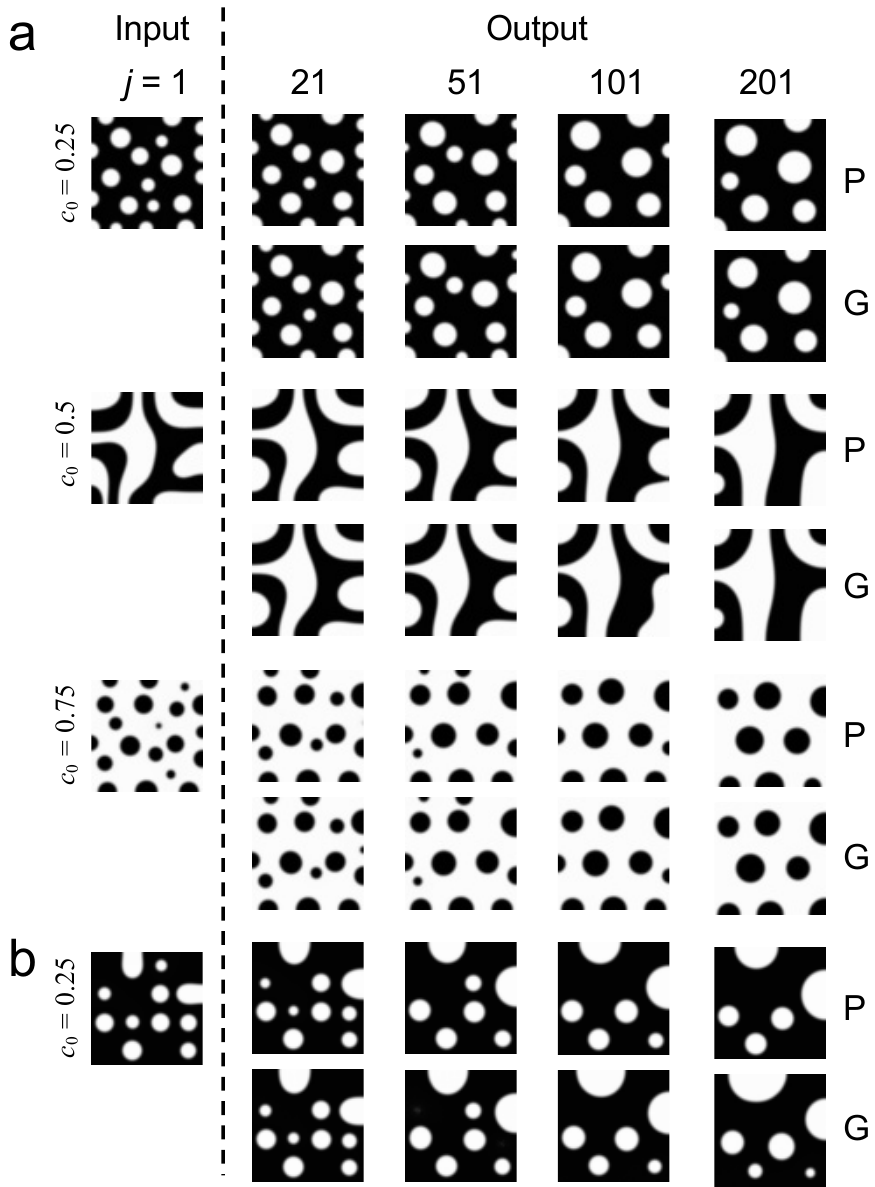}
	\caption{ \textbf{Application of RNN to predicting spinodal decomposition.} 
		\textbf{a.} Comparison between predictions (P) and ground truth (G) from two testing cases, in which RNN outputs 200 frames based on 1 input frame of spinodal structure generated from random perturbation to a system of uniform composition. 
		\textbf{b.} RNN prediction of the evolution of an artificial bi-phasic configuration, in which second-phase particles ($c=1$) of randomly chosen radii are orderly arranged within the primary phase ($c=0$). 
	}
	\label{fig:spinodal1}
\end{figure}

\begin{figure}[!thp]
	\vskip 0.15in
	\centering
	\includegraphics[width=0.9\columnwidth]{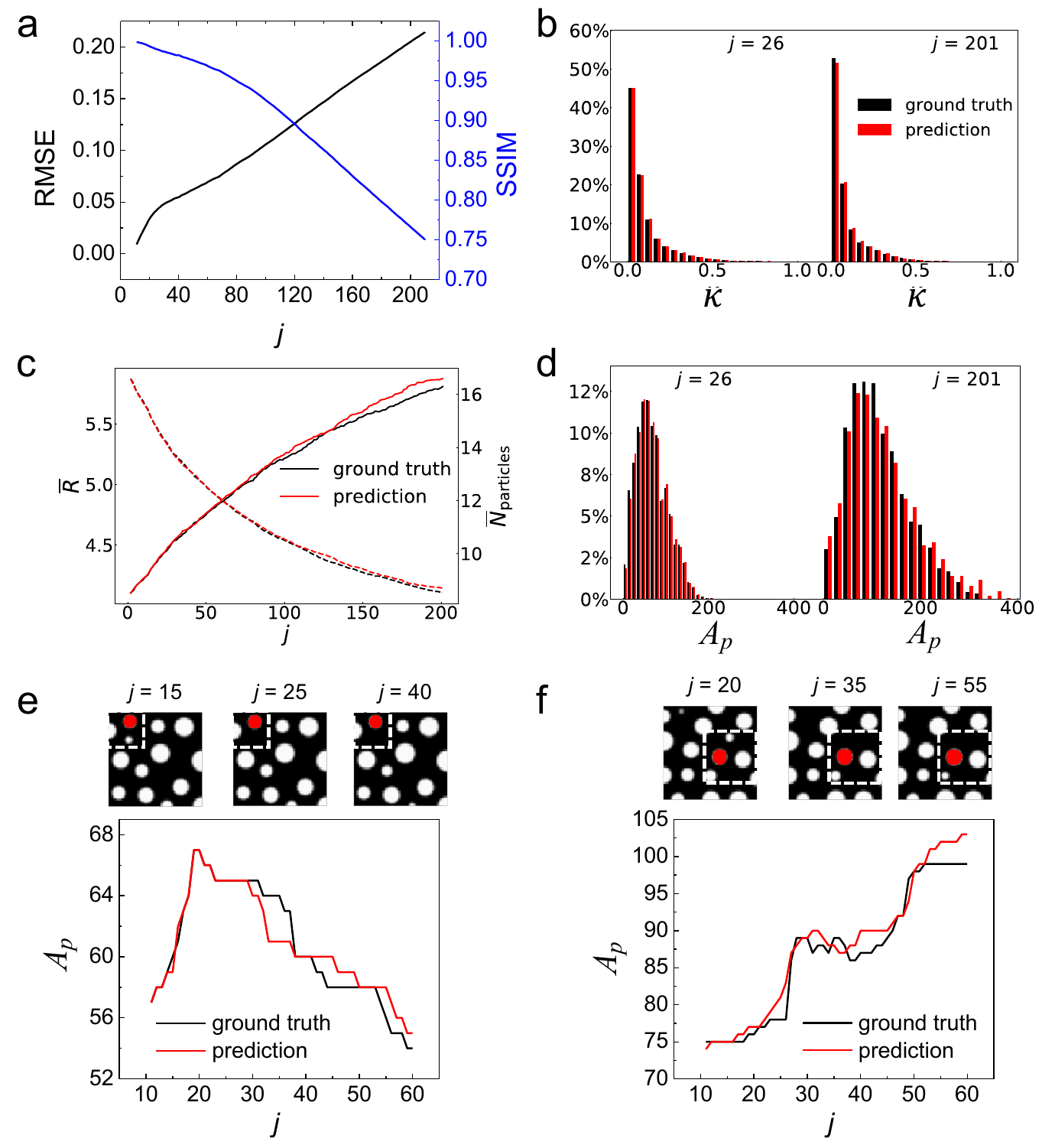}
	\caption{ \textbf{Accuracy of RNN in predicting spinodal decomposition.}
		\textbf{a.} RMSE and SSIM of RNN predictions averaged over 510 testing cases as a function of the frame index $j$. 
		\textbf{b.} Distribution of the interface segment curvature $\kappa$ at $j = $ 26 and 201 in 170 testing cases with $c_0$ = 0.5.
		\textbf{c. and d.} Evolution of the average second-phase particle radius $\langle R\rangle$ (\textbf{c}) and particle area $A_p$ distribution (\textbf{d}) in 170 testing cases with $c_0$ = 0.25. $\langle R\rangle$ is calculated as $\sqrt{\langle A_p\rangle/\pi}$. 
		\textbf{e. and f.} (top) Examples of local morphological evolution predicted by RNN from two testing cases with $c_0$ = 0.25. (bottom) Size evolution of the red particle in the images as predicted by RNN vs ground truth. 
	}
	\label{fig:spinodal2}
\end{figure}

We perform temporal extrapolation tests on the trained model in a similar way to the case of grain growth.
RNN is asked to output 200 frames, or 10 times of the training clip length, given one input frame that is 
taken from the 50th frame of a simulation starting from a uniform mixture.
75\% of the output frames ($j$ = 51 -- 200) thus fall outside the time span of the training sets.
In addition, predictions based on 10 input frames are also tested.
The results are presented in Supplementary Figure S3, which show similar performance as those with 1 input frame only,
which indicates that the information contained in the initial system configuration is sufficient for RNN to correctly project the evolution trajectory. 
\textbf{Figure} \ref{fig:spinodal1}a showcases several examples from a total of 510 tests with 170 each having $c_0$ = 0.25, 0.5 or 0.75.
The short-term predictions up to $j\sim$ 50 closely resemble the ground truth,  
which is quantified by the low RMSE ($< 0.06$) and high SSIM ($> 0.97$) in \textbf{Figure} \ref{fig:spinodal2}a.
While the discrepancy gradually accumulates with time and visible difference appears at the later stage, 
the long-term predictions are realistic looking and no artifacts can be discerned. 
In addition, Supplementary Figure S4 shows that RNN well conserves the mass in the system, 
with the average concentration differing less than 6\% from $c_0$ after 200 output frames. 
Morphology-wise, it is difficult to tell by human eyes whether the images are generated by RNN or simulations.
Such similarity is corroborated by the statistical analysis of the microstructure.
In \textbf{Figure} \ref{fig:spinodal2}b, we compare the interface curvature distributions in the predicted versus ground truth images of 170 testing cases with $c_0$ = 0.5,
which have a bicontinuous two-phase morphology. 
The agreement is very good in both short and long terms,
which can be quantified by the Euclidean distance between the two distributions:
0.0028 at frame $j$ = 26 and 0.014 at $j$ = 201. 
On the other hand, systems with $c_0$ = 0.25 or 0.75 contain individual particles of the minority phase ($c$ = 1 or 0) dispersed within the majority phase. 
\textbf{Figure} \ref{fig:spinodal2}c shows the time dependence of the average particle size $\langle R\rangle$ for 170 tests with $c_0$ = 0.25. 
The corresponding particle size distributions are presented in \textbf{Figure} \ref{fig:spinodal2}d. 
The comparison is again satisfactory. 
The predicted $\langle R\rangle$ has a maximal error of 1.89\% within the test period,
and the Euclidean distance between the predicted and true size distributions is only 0.01 at $j$ = 26 and 0.034 at $j$ =  201. 

We next perform the spatial extrapolation tests by applying the trained model to a larger 256$\times$256-pixel domain.
As shown in Supplementary Figure S5, RNN performs equally well in the extended system with comparable RMSE and SSIM as in the smaller domain.  
Furthermore, \textbf{Figure} \ref{fig:spinodal1}b shows an example which tests the ability of RNN in predicting the evolution of configurations ``foreign'' to the training datasets. 
The initial configuration in the test is created by placing circular particles of $c=1$ with random radii on square lattice sites in the matrix of $c=0$. 
Though never seeing such a configuration during training, RNN captures its evolution very well.  

The impressive extrapolation capability of RNN when applied to spinodal decomposition also implies its understanding of the physical rules of this phenomenon.
The coarsening of the spinodal structure is thermodynamically driven by the interface curvature dependence of chemical potentials (i.e. Gibbs-Thomson effect) and kinetically limited by the species diffusion. 
\textbf{Figure} \ref{fig:spinodal2}b and d show that RNN grasps the Gibbs-Thomson effect, which causes the fraction of low-curvature interface segments to increase with time, 
and \textbf{Figure} \ref{fig:spinodal2}c confirms that the diffusion-controlled coarsening kinetics is captured by the model.
Apart from the accurate statistical representation, the examples in \textbf{Figure} \ref{fig:spinodal2}d and e illustrate that RNN is also capable of predicting subtle local morphological changes.  
The fate of the particle highlighted by red in \textbf{Figure} \ref{fig:spinodal2}e is determined by the relative sizes of its neighbor particles,
which exchange mass between each other via diffusion due to the size-dependent chemical potential.  
The red particle first grows at the expense of a smaller neighbor, but subsequently shrinks by losing mass to the other two bigger particles nearby. 
In \textbf{Figure} \ref{fig:spinodal2}f, the particle in red receives an incoming diffusion flux from two smaller adjacent particles.
Its growth rate exhibits two bursts, which coincide with the complete dissolution of the two particles. 
RNN's ability to predict detailed evolution features as demonstrated in these examples further inspires confidence in its comprehension of the underlying physics.

\subsection{Dendrite growth}
In the last example, we give RNN a more challenging task to predict dendritic crystallization patterns. 
During crystal growth, dendritic structures like the beautiful snowflakes often form due to the morphological instability of the growth front,
which is promoted by the negative temperature and/or species concentration gradient(s) ahead of the phase boundary and the interface energy anisotropy.  
Such instability phenomena are intrinsically difficult to predict.  
In addition, dendrite growth is a multiphysical process coupling phase transformation, long-range mass and heat transport and interface instability.
As a result, microstructure images fed to RNN do not contain the complete information of the system state,
which further increases the difficulty of making accurate predictions. 
\begin{figure}[!thp]
	\centering
	\includegraphics[width=0.8\columnwidth]{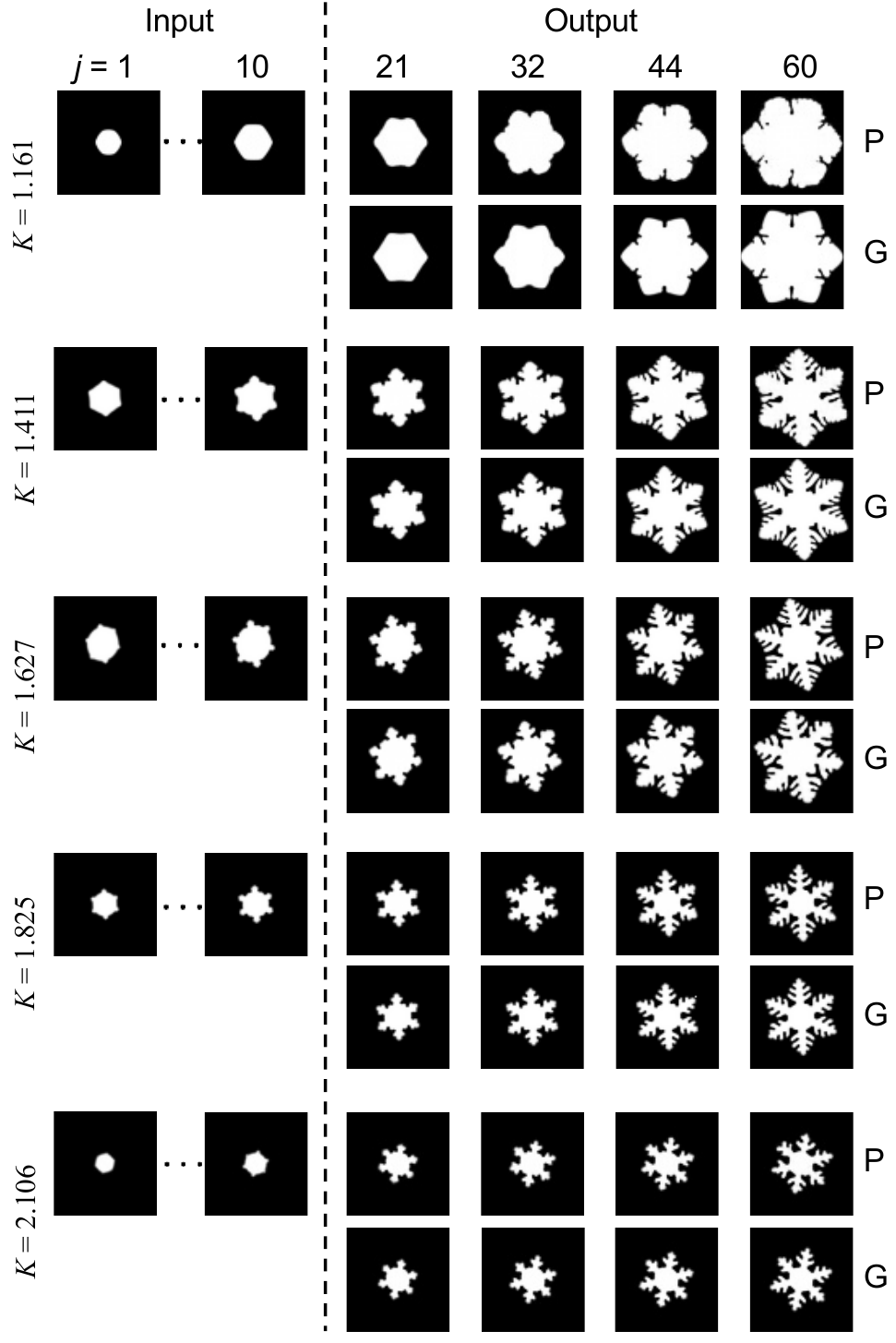}
	\caption{ \textbf{Application of RNN to predicting dendritic crystal growth.} RNN predictions (P) vs. ground truth (G) from five testing cases with different $K$ values, in which RNN outputs 50 frames based on 10 input frames.
	}
	\label{fig:dendrite1}
\end{figure}

Here we generate training data using a phase-field model of solidification in pure systems by Kobayashi\cite{kobayashi1993modeling}. 
As described in the Methods section, the spatiotemporal evolution of the system state is described by two coupled PDEs for the temperature ($T$) and phase-field ($\phi$) variables. 
$\phi$ distinguishes between the solid ($\phi$=1) and liquid ($\phi$=0) phases during solidification. 
We use $\phi(t,x,y)$ to create the microstructure images.
$T$ and other parameters in the governing equation (Eq.~\ref{eq:dendritepfm2}) such as the normalized latent heat $K$ are thus hidden to the learning process.
We perform phase-field simulations on a 64$\times$64 mesh, in which a small solid nucleus is placed at or near the center and surrounded by the supercooled liquid phase. 
The training and validation sets contain 800 and 200 simulations, respectively.
To enrich the training data, each simulation has a different nucleus, crystal orientation $\theta_0$ and $K$.
Specifically, $K$ is randomly chosen from (1.2, 2) and $\theta_0$ from (0, $\pi/3$) (crystal is assumed to have six-fold symmetry).
The nucleus is given random shape (circle, rectangle or ellipse), size (2 -- 6 pixels) and off-center distance ($\pm$5 pixels in $x$ and $y$ directions).
Similar to the case of spinodal decomposition, 100 image frames with equal time interval are obtained from a simulation and divided into eight staggered 20-frame training clips.

\begin{figure}[!thp]
	\centering
	\includegraphics[width=1\columnwidth]{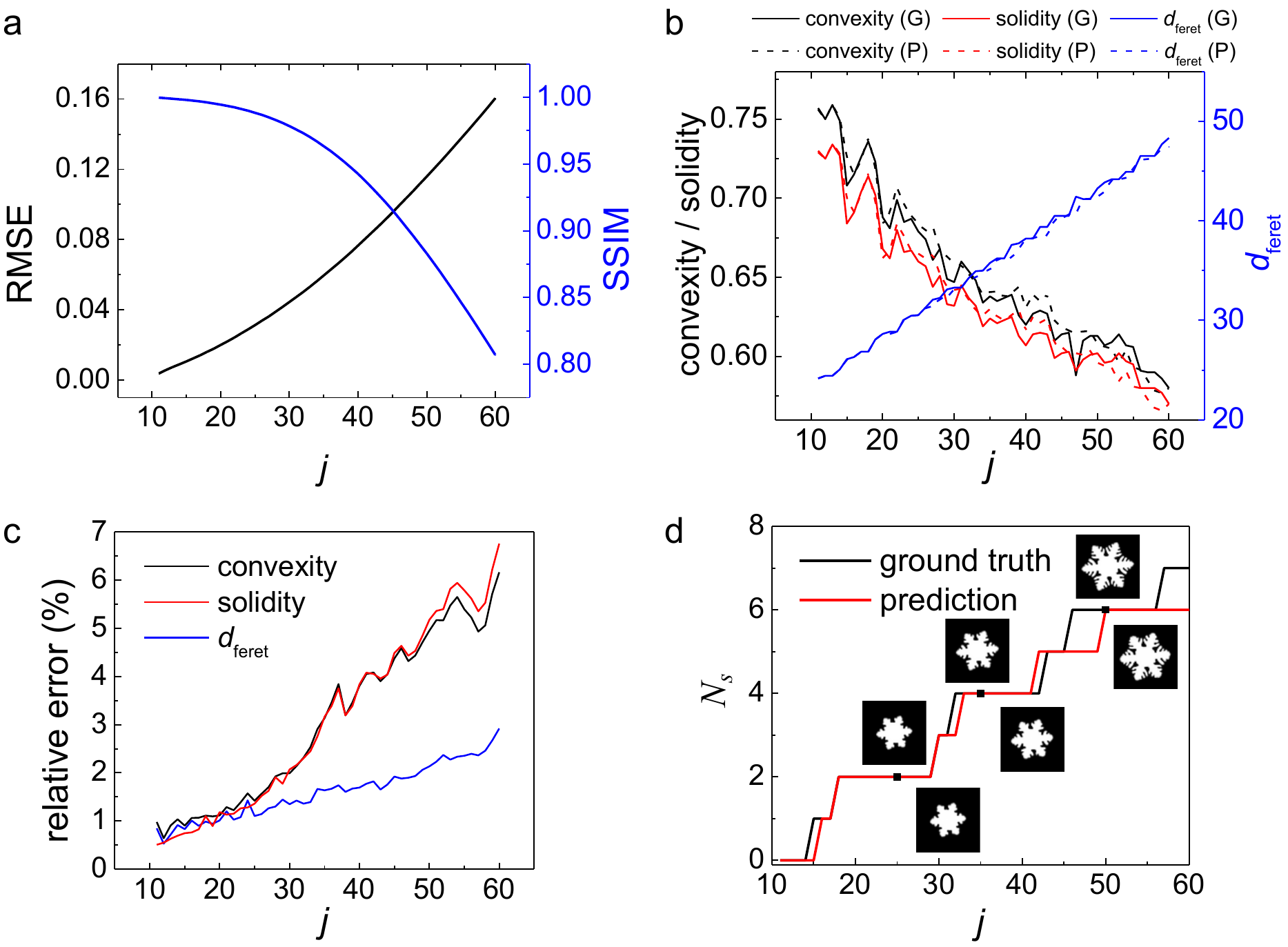}
	\caption{ \textbf{Accuracy of RNN in predicting dendritic crystal growth.}
		\textbf{a.} RMSE (black) and SSIM (blue) of the predictions averaged over 600 testing cases. 
		\textbf{b.} Time evolution of the Feret diameter $d_{feret}$, convexity and solidity of a growing crystal from a testing case. Solid lines are ground truth (G), and dashed lines are predictions (P). The shape descriptors are calculated in imageJ after image binarization. Convexity is defined as $L_h/L_c$, where $L_c$ is the crystal perimeter and $L_h$ is the perimeter of the convex hull of the crystal. Solidity is defined as $A_c/A_h$, where $A_c$ is the crystal area and $A_h$ is the area surrounded by the convex hull of the crystal shape. 
		\textbf{c.} Relative errors of predicted $d_{feret}$, convexity and solidity of crystals averaged over 600 testing cases as a function of image frame index $j$.
		\textbf{d.} Development of secondary branches on the dendritic crystal in a testing case. $N_{s}$ is the number of secondary branches on a primary branch of the dendrite. Insets above and below the curves show the crystal shape from the ground truth and predictions, respectively, at times marked by the black squares. 
	}
	\label{fig:dendrite2}
\end{figure}

In testing, the trained RNN model is required to predict 50 frames from 10 consecutive input frames, which are taken from the first half of a simulation.
Predictions are not extended to longer time because the dendrite tips already approach the domain boundaries after 50 output frames in many tests and growth stagnates subsequently. 
Instead, we focus on conducting the extrapolation tests in the model parameter space. 
Specifically, $K$ is randomly and uniformly selected from (0.8, 2.4) to generate ground truth data in the testing cases.
This means that half of the selected $K$ values fall outside its range in the training data, which is (0.8, 2).
$\theta_0$ and the solid nucleus shape are also randomized.
\textbf{Figure} \ref{fig:dendrite1} presents several examples from a total of 600 testing cases. 
The predicted dendritic structure matches the ground truth well in all the cases
even at $K$ = 1.161 and 2.106, 
which are outside the scope of training data.
In particular, RNN captures the fine features of the dendrites such as the locations of secondary side branches.
It can be seen that the crystal growth pattern depends strongly on $K$. 
Smaller $K$ results in thicker primary branches and more compact morphology.
RNN manages to recognize the correct evolution trajectory based on the input images without prior knowledge of the underlying $K$ value.
\textbf{Figure} \ref{fig:dendrite2}a shows RMSE and SSIM of the predictions averaged over all of the 600 testing cases.
RNN fares well in pixel-wise comparisons although the prediction error increases faster with time than in the cases of grain growth and spinodal decomposition,
which can be attributed to the more complex physics of the dendrite growth process.

As a more revealing indicator of RNN's performance, we use several shape descriptors (Feret diameter $d_{feret}$, convexity, solidity) to characterize the dendrite morphology.  Feret diameter, which is defined as the maximum distance between two parallel tangent lines touching the shape, provides a measure of the linear dendrite dimension.
Convexity and solidity quantify the degrees of concavity and compactness of the crystal. 
\textbf{Figure} \ref{fig:dendrite2}b shows the time evolution of these descriptors from one test while their average errors for all the 600 tests are plotted in \textbf{Figure} \ref{fig:dendrite2}c.
It can be seen that RNN accurately predicts the dendritic shape evolution with the average error less than 7\% throughout the tests.
In addition to global metrics, we also examine how well RNN reproduces local dendritic structural features. 
In \textbf{Figure} \ref{fig:dendrite2}d, the number of secondary branches formed on a primary branch in a test is plotted as a function of time.
It shows that RNN performs very well in predicting the frequency of the side-branching events occurring near the dendrite tip.   

\section*{Discussion}
In addition to the prediction accuracy, we compare the computational efficiency of using RNN for microstructure evolution predictions with that of PDE-based simulations. 
During training and testing, the time interval between two RNN output frames is 80 times that of the average time step size used in the grain growth simulation, 370 times in spinodal decomposition and 7 times in dendritic crystal growth.
These large time spacings will quickly render PDE solvers unstable. 
This illustrates a significant computational benefit of RNN, that its performance is not limited by the numerical stability of PDEs and can thus make reliable predictions at much larger time step size. 
In the grain growth example, RNN's advantage in spatial coarsening is also demonstrated.
Because of the diffuse-interface representation, a grain boundary needs to be resolved by at least 5--6 pixels in phase-field simulations to maintain desired numerical accuracy\cite{Moelans2008C268}. 
However, RNN is not subject to the same spatial resolution requirement and can predict system evolution on a coarser mesh (64$\times$64) than used in phase-field simulations (256$\times$256). 
Accordingly, the improved efficiency in time marching and spatial coarse-graining results in computational saving. 
We benchmark E3D-LSTM's running time on a compute node with 4 NVidia GeForce GTX 1080-TI GPUs. 
After spending 130 -- 450 seconds to initialize and load pre-trained models, it takes on average 2.1 s to predict a case for grain growth (200 64$\times$64-pixel output frames), 3.8 s for spinodal decomposition (200 64$\times$64-pixel output frames), and 0.56 s for dendrite growth (50 64$\times$64-pixel output frames).
In comparison, numerical simulations of the three examples, which are implemented by an in-house C code (grain growth) and COMSOL Multiphysics 5.3 (spinodal decomposition, dendrite growth) on a desktop with an Intel i7 3.2GHz CPU, require an average running time of 1551 s for grain growth, 350 s for spinodal decomposition and 44 s for dendrite growth. 
The comparison shows that RNN is computationally efficient, especially when applied to a large number of cases so that the overhead associated with initialization is small. Note that the speedup estimation made here is conservative: the PDEs employed were simple textbook examples and substantially cheaper than more sophisticated equations in real applications, and were implemented in efficient C codes, while the neural nets were based on research Python codes and not optimized for physics simulation.

The overall efficiency of RNN in predicting microstructure evolution also depends on the training data size and the efforts and resources required for data collection. 
Supplementary Figure S6 shows the dependence of validation error on the number of training clips $N_{clip}$ in the cases of plane wave propagation and grain growth.
In both cases, the improvement in model performance becomes negligible after $N_{clip}$ goes beyond $\sim$2000.
On the other hand, we find that increasing the length of training clips beyond 20 frames does not significantly improve the prediction accuracy. 
For all of the examples in this work, the time spent on generating the training datasets is comparable to the model training time. 
Therefore, the data requirement of RNN should not present a major obstacle to its applications. 

Despite the overall very impressive performance, our tests show that the learning rate and predictive power of RNN vary with the nature of the microstructure evolution phenomena it is applied to. 
Among all the examples, RNN demonstrates the best learning ability in predicting grain growth because its evolution rules are localized,
which could be relatively easily recognized by E3D-LSTM through 3D convolution operations that specialize in remembering local, short-term motion.
In contrast, training RNN to predict spinodal decomposition is more challenging because the long-range mass transport inherent in the process creates longer and stronger spatiotemporal correlation,
which requires more convolution operations and long-term memory states to extract the essential features.
In fact, the model can be successfully trained to predict grain growth with only two E3D-LSTM layers, 
but 4 layers are needed for spinodal decomposition to reach similar performance.   
Compared to grain growth, we also find it necessary to include longer image sequences (100 frames) into the training datasets for spinodal decomposition to better inform RNN of the evolution trajectories and achieve comparable prediction accuracy. 
Predicting dendrite growth presents additional challenges due to the interface instability and the existence of hidden variable ($T$) not directly seen by RNN.
However, potential improvement could be achieved by encoding both $\phi(t,x,y)$ and $T(t,x,y)$ into multi-channel images to let RNN learn the evolution of not only the microstructure morphology but also other relevant fields, 
which could be a general strategy to effectively predict microstructure evolution governed by complex multiphysics principles.  

In summary, we train a convolutional recurrent neural network (E3D-LSTM) to predict the spatiotemporal evolution of materials microstructure. 
Using training data from four distinct evolution processes (plane wave propagation, grain growth, spinodal decomposition and dendritic crystal growth), the RNN, which is composed of the same network architecture and agnostic about the underlying physics, was able to adapt efficiently to different evolution rules.
The ability of RNN to generalize learning beyond the training datasets is systematically examined by a series of extrapolation tests, even though the neural net is ,
In addition to performing very well in piecewise comparison with ground truth in short-term predictions, 
RNN accurately describes the statistical properties of microstructures over long periods up to ten folds of the training data's time span.
Without additional training, neural nets trained on small-size images can be straightforwardly applied to larger systems with comparable accuracy. 
The method can reliably predict the evolution of microstructures whose morphology or underlying materials parameters differ qualitatively from the training data.   
The spatiotemporal, configurational and parametric extensibility demonstrated by RNN suggests that it is capable of learning the evolution rules of the microstructure phenomena considered here, 
which provides the physical basis for its practical applications.
Computationally, RNN is not restricted by the numerical stability of PDE solvers and can employ time step size 1-2 orders of magnitude larger than PDE-based simulations in our tests. 
The ML approach demonstrated in this study provides a valuable complement and/or ßalternative to physics-based simulations for predicting microstructure evolution, which could be especially attractive in situations where there exist unknown materials parameters or evolution principles are not fully understood.

\section*{Methods}
\subsection*{I. Recurrent Neural Network}
Unlike static data without temporal context, sequential data such as the microstructure evolution trajectories in the form of image sequences require special treatment for deep neural networks to learn efficiently and accurately. 
Designed to take advantage of the temporal information of sequential inputs, RNN along with its LSTM variants were first successfully employed in voice recognition and natural language processing.
Recently, Shi et al.\cite{xingjian2015convolutional} proposed a convolutional LSTM model for image sequence prediction, which uses CNN instead of fully connected layers in vanilla RNN for latent-feature extraction, 
and combines it with LSTM for learing time evolution to make full use of features in both spatial and temporal domains. 
More recent studies replace the initially stacked chain structure\cite{xingjian2015convolutional} with sophisticated neural nets to improve information flow and reach better performance. 

For example, Yunbo Wang and co-workers developed a series of neural networks for spatiotemporal predictive learning\cite{Wang2017PredRNN, Wang2018PredRNN++, Wang2019E3D-LSTM}. The latest Eidetic 3D LSTM (E3D-LSTM) model is employed in our study. Compared with other state-of-the-art models that use 2D convolution operations, E3D-LSTM integrates 3D (one temporal and two spatial dimensions) convolution (3D-Conv) deep into RNNs, which is effective for modeling local representations in a consecutive manner. As shown in Figure 1(c) of Ref. \cite{Wang2019E3D-LSTM}, successive input frames are encoded by 3D-Conv encoders before being fed to E3D-LSTM units. 
Outputs of E3D-LSTM units are decoded with a 3D-Conv layer to obtain the real-space predictions. 
Besides adopting 3D-Conv as its basic operations, E3D-LSTM exploits a self-attention mechanism to memorize long-term interactions in addition to short-term motions. 
This is achieved by implementing two distinct memory states in E3D-LSTM: spatiotemporal memory and eidetic 3D memory. 
The former is designed to capture the short-term motion\cite{Wang2017PredRNN} 
while the latter computes the relation between local patterns and the whole memory space to distinguish and revoke temporally distant memories. 

\textbf{\textit{Model setup}}: 
Each data point in the training sets is a sequence of $N_t$ 2D images generated by a scalar field $c(t,x,y)$ ( $0\leq c \leq 1$, $t=1$\ldots$N_t$, $x=1$\ldots$N_x$, $y=1$\ldots$N_y$). The spatial dimensions $N_x$ and $N_y$ are 64 unless otherwise stated.
For each problem considered, the training dataset is a 4D array $c_i(t, x, y)$ with $N_{total}$ image sequences ($i= 1$\ldots$N_\text{total}$).
Following Ref.~\cite{Wang2019E3D-LSTM}, four E3D-LSTM layers are stacked together in the model (only two layers in the case of grain growth), each with 64 hidden features.
The model is implemented in Tensor Flow\cite{abadi2016tensorflow} and trained on 4 NVidia V100 or 1080-TI GPUs.
Typical training time is 36--48 hours, with an initial learning rate of $10^{-3}$ that gradually decays to $10^{-5}$.

\textbf{\textit{Data augmentation}}: 
Training data are augmented by performing symmetry operations of the 2D point group $4mm$ on the original images,
which transform $c(t,x,y)$ to $c(t, \bar{x}, y)$, $c(t, x, \bar{y})$, $c(t, \bar{x}, \bar{y})$, $c(t, y, x)$, $c(t, \bar{y}, x)$, $c(t, y, \bar{x})$ and $c(t, \bar{y}, \bar{x})$ ($\bar{x} \equiv N_x+1-x$, $\bar{y} \equiv N_y+1-y$). 
Such data transformations can be achieved by array rearrangements and do not require additional float-point calculations. 

\textbf{\textit{Analysis methods}}: 
RMSE and SSIM are used in pixel-wise comparison between ground truth and predictions.
RMSE is defined as 
\begin{equation}
\text{RMSE} = \sqrt{\sum_{i=1}^{N_x}\sum_{j=1}^{N_y}\frac{(p_\text{g}(i,j) - p_\text{p}(i,j))^2}{N_x N_y}}
\end{equation}
where $p_\text{g}(i,j)$ and $p_\text{p}(i,j)$ are the pixel values of ground truth and predictions, respectively.
SSIM\cite{wang2004image} is defined as 
\begin{equation}
    \text{SSIM} = \frac{(2\overline{p}_\text{g} \overline{p}_\text{p} + c_1) (2\sigma_\text{gp}+c_2)}{(\overline{p}_\text{g}^2 + \overline{p}_\text{p}^2 + c_1)(\sigma_\text{g}^2 +\sigma_\text{p}^2 + c_2 )}
\end{equation}
where $\overline{p}_\text{k}$ and $\sigma_\text{k}$ ($\text{k} = \text{g, p}$) are the average pixel value and variance of ground truth or predictions, respectively, and $\sigma_\text{gp}$ is their covariance. $c_1$ and $c_2$ are small constants and chosen to be $c_1= (0.01L)^2$ and $c_2=(0.03L)^2$, where $L$ is the range of pixel values.
The Euclidean distance between the distributions of quantity $q$ from RNN predictions and ground truth is defined as 
\begin{equation}
d = \sqrt{\sum_{i=1}^n (q_\text{g}^i - q_\text{p}^i)^2}
\end{equation}
where $n$ is the number of bins within the interval between the minimum and maximum of $q$, and $q^i_\text{g}$ and $q^i_\text{p}$ are normalized counts in the $i$-th bin of the ground truth and predictions, respectively. $n=20$ is used for all the calculations. 

\subsection*{II. Simulation method}
Phase-field simulations are employed to generate the ground truth for three microstructure evolution processes, i.e. grain growth, spinodal decomposition and dendritic crystal growth. 
Phase-field method is a powerful computational technique for modeling microstructure evolution in diverse materials systems\cite{Chen2002PFMreview, Moelans2008C268, steinbach2009phase}. 
In a phase-field model, different phases are represented by one or multiple order parameters, and their interfaces are tracked by the level sets of the order parameters. 
Spatiotemperoal evolution of the microstructure is described by the governing equations of the order parameters derived from thermodynamic and kinetic principles.

\textbf{\textit{Grain growth}}:
Isotropic grain growth in 2D polycrystalline structure is simulated by a multi-order-parameter phase-field model\cite{Moelans2008PRB}. 
In the model, a set of order parameters $\{\eta_1(x), \eta_2(x), ..., \eta_N(x)\}$ are used to represent $N$ distinct grain orientations. The free energy of the system is expressed as
\begin{equation}\label{eq:gg1}
	F = \int \left[ f(\eta_1, \eta_2, ..., \eta_N) + \frac{\nu}{2} \sum^N_{i=1}\left( \nabla\eta_i \right)^2 \right] dV
\end{equation}
where the homogeneous free energy density $f$ is given by
\begin{equation}\label{eq:gg2}
	f = m \left[ \sum^N_{i=1} \left( \frac{\eta^4_i}{4} - \frac{\eta^2_i}{2} \right) + \frac{3}{2} \sum^N_{i=1}\sum^N_{j>i} \eta^2_i \eta^2_j + \frac{1}{4}  \right]
\end{equation}
which has $N$ local minima located at $(\eta_1, \eta_2, ..., \eta_N) = (1, 0, ..., 0), (0, 1, ..., 0), ..., (0,0,...,1)$.  
The evolution of ${\eta_i(x)}$ ($i = 1\ldots N$) follows the time-dependent Ginzburg-Landau or Allen-Cahn\cite{allen1972ground,allen1973correction} equation
\begin{equation}\label{eq:gg3}
	\frac{\partial\eta_i}{\partial t} = -L\frac{\delta F}{\delta \eta_i}
\end{equation}
In all the simulations, dimensionless parameters $N=100$, $m= 1$, $\nu=1$ and $L=1$ are used.
The initial polycrystalline structure is generated by Vornoi tessellation\cite{aurenhammer2000voronoi} with 100 grains. 
Eq.~\ref{eq:gg3} is solved by the forward Euler finite difference scheme with periodic boundary conditions and grid spacing $\Delta x$ = 1 and time step size $\Delta t$ = 0.2.
Single-channel images of the polycrystalline structure are generated by assigning $\sum^{N}_{i=1} \eta^3_i$ as the pixel value so that pixels are close to 0 in the grain boundary region and 1 inside grains. 

\textbf{\textit{Spinodal decomposition}}:
Spinodal decomposition is simulated by the Cahn-Hilliard equation\cite{cahn1958free},
\begin{equation}\label{eq:cahn1}
    \frac{\partial c}{\partial t}=\nabla\cdot \left[ Mc(1-c)\nabla \left( \frac{\partial f_{chem}}{\partial c}-\epsilon\nabla^2c \right) \right],
\end{equation}
where $c$ is the molar fraction of a species in a binary system. 
We use the regular solution model to describe the homogeneous free energy density:  
\begin{equation}\label{eq:cahn2}
    f_{chem}(c)=RT[c\ln c+(1-c)\ln(1-c)]+\omega c(1-c)
\end{equation}
with a positive value is assigned to the regular solution coefficient $\omega$ to favor phase separation. 
Eqs. \eqref{eq:cahn1}-\eqref{eq:cahn2} are solved with no-flux boundary conditions. 
Dimensionless parameter values $\omega = 0.27397$, $\epsilon = 0.1682$ and $M = 1$ and mesh spacing $\Delta x = 1$ are used in all of the simulations. 
Eq.~\ref{eq:cahn1} is solved with an implicit backward differentiation formula (BDF) solver in COMSOL Multiphysics with an average dimensionless time step size of 4.01.
Images are output from simulations at a time interval of 1500, or an average of 370 steps between two frames.

\textbf{\textit{Dendrite growth}}:
We use a phase-field model developed by Kobayashi\cite{kobayashi1993modeling} to simulate the dendritic solidification process in a pure material system. 
Compared to other more quantitative models\cite{karma1998quantitative, steinbach2009phase},
this model is chosen for its simplicity since the purpose of this work is not to study dendritic growth but use it as an example to evaluate RNN. 
The system state is described by the temperature field $T$ and an order parameter $\phi$, which distinguishes between the solid ($\phi = 1$) and liquid ($\phi = 0$) phases.
The free energy of the system is given by 
\begin{equation}
    F[\phi,T]=\int \left[ \frac{1}{2}\epsilon(\theta)^2|\nabla\phi|^2+f(\phi,T)\right]\mathrm{d}\mathbf{r},
\end{equation}
where the anisotropy of the solid/liquid interface energy is controlled by the orientation dependence of the gradient energy coefficient:
$\epsilon(\theta)={\epsilon_0}(1+\delta\cos[n(\theta-\theta_0)])$, 
where $\theta$ represents the interface normal and is calculated from the gradient of $\phi$ as $\theta=arctan(-\phi_y/\phi_x)$.
We employ $n=6$ in simulations to produce dendrites with sixfold symmetry. 
$f$ is a double-well potential
\begin{align}
    f(\phi,T) & =  \frac{1}{4}\phi^4-\left[\frac{1}{2}-\frac{1}{3}m(T)\right]\phi^3+\left[\frac{1}{4}-\frac{1}{2}m(T)\right]\phi^2,\\
    m(T) & = \frac{\alpha}{\pi}\arctan[\gamma(T_{\mathrm{eq}}-T)].
\end{align}
where $T_{\mathrm{eq}}$ is the solid/liquid equilibrium temperature.
The time evolution of the coupled $\phi$ and $T$ fields is governed by 
\begin{align}
    \label{eq:dendritepfm1}
    \tau \frac{\partial \phi}{\partial t} & =-\frac{\delta F}{\delta \phi}, \\
    \label{eq:dendritepfm2}
    \frac{\partial T}{\partial t} & =\nabla^2 T  + K \frac{\partial \phi}{\partial t}
\end{align}
where constant $K$ represents the latent heat. 
The following dimensionless parameters are used in all the simulations: $\alpha=0.9$, $\gamma=10$, $T_\mathrm{eq}=1$,  $\tau=0.001$, $\epsilon_0=0.01$, $\delta=0.03$ while $K$ and $\theta_0$ are varied.
The system has a uniform initial temperature at $T(t=0,x,y) = 0$.
Eq.~\ref{eq:dendritepfm1} and \ref{eq:dendritepfm2} are solved with a BDF solver in COMSOL Multiphysics with mesh spacing $\Delta x = 1$ and average time step size $\Delta t = 5.7\times 10^{-4}$. 
Images are output from simulations at a time interval of 0.004, or an average of 7 time steps between two frames.

\section*{Data availability} 
The data that support the findings of this study are available from the corresponding authors upon request.

\section*{Acknowledgements}
K.Y., Y.C and M.T.\ acknowledge support from DOE under project number DE-SC0019111. 
Y.Z. acknowledges support from NSF under project number CMMI-1929949.
The work of D.A., B.S.\ and F.Z.\ was supported by the Critical Materials Institute, an Energy Innovation Hub funded by the U.S. Department of Energy, Office of Energy Efficiency and Renewable Energy, Advanced Manufacturing Office, and performed under the auspices of the U.S. Department of Energy by LLNL under Contract DE-AC52-07NA27344. 
Phase-field simulations were performed on supercomputers at the Texas Advanced Computing Center (TACC) at The University of Texas.
RNN training and testing were performed on supercomputers at LLNL and TACC.

\section*{Author contributions}
F.Z. conceived the project. F.Z. and M.T. supervised the project. 
K.Y., Y.Z. and Y.C. performed phase-field simulations. 
K.Y., Y.C. and F.Z. performed RNN training and testing. 
K.Y., Y.C., Y.Z., M.T., D.A., B.S. and F.Z. analyzed and discussed the results.
M.T., F.Z., K.Y. and Y.C. wrote the manuscript with inputs from other authors.

\bibliographystyle{Science}

\bibliography{paper_0812_KY}

\end{document}


\title{Supplementary Information for\\
	Self-supervised Learning and Prediction of Microstructure Evolution with Recurrent Neural Networks}
	\author{Kaiqi Yang}
	\author{Yifan Cao}
	\author{Youtian Zhang}
	\author{Ming Tang}
	\email{mt20@rice.edu}
	\affiliation{Department of Materials Science and NanoEngineering, Rice University, Houston, TX 77005, USA}
	\author{Daniel Aberg} 	
	\author{Babak Sadigh}	
	\author{Fei Zhou} 
	\email{zhou6@llnl.gov}
	\affiliation{Physical and Life Sciences Directorate, Lawrence Livermore National Laboratory, Livermore, CA 94550, USA} 
	
	\maketitle

\newcommand{\beginsupplement}{%
	\renewcommand{\figurename}{\textbf{Supplementary Figure}}
	\renewcommand{\tablename}{Supplementary Table}
        \setcounter{table}{0}
        \renewcommand{\thetable}{\textbf{S\arabic{table}}}%
        \setcounter{figure}{0}
        \renewcommand{\thefigure}{\textbf{S\arabic{figure}}}%
     }
     
\beginsupplement

\begin{figure}[!thp]
	\centering
	\includegraphics[width=0.8\columnwidth]{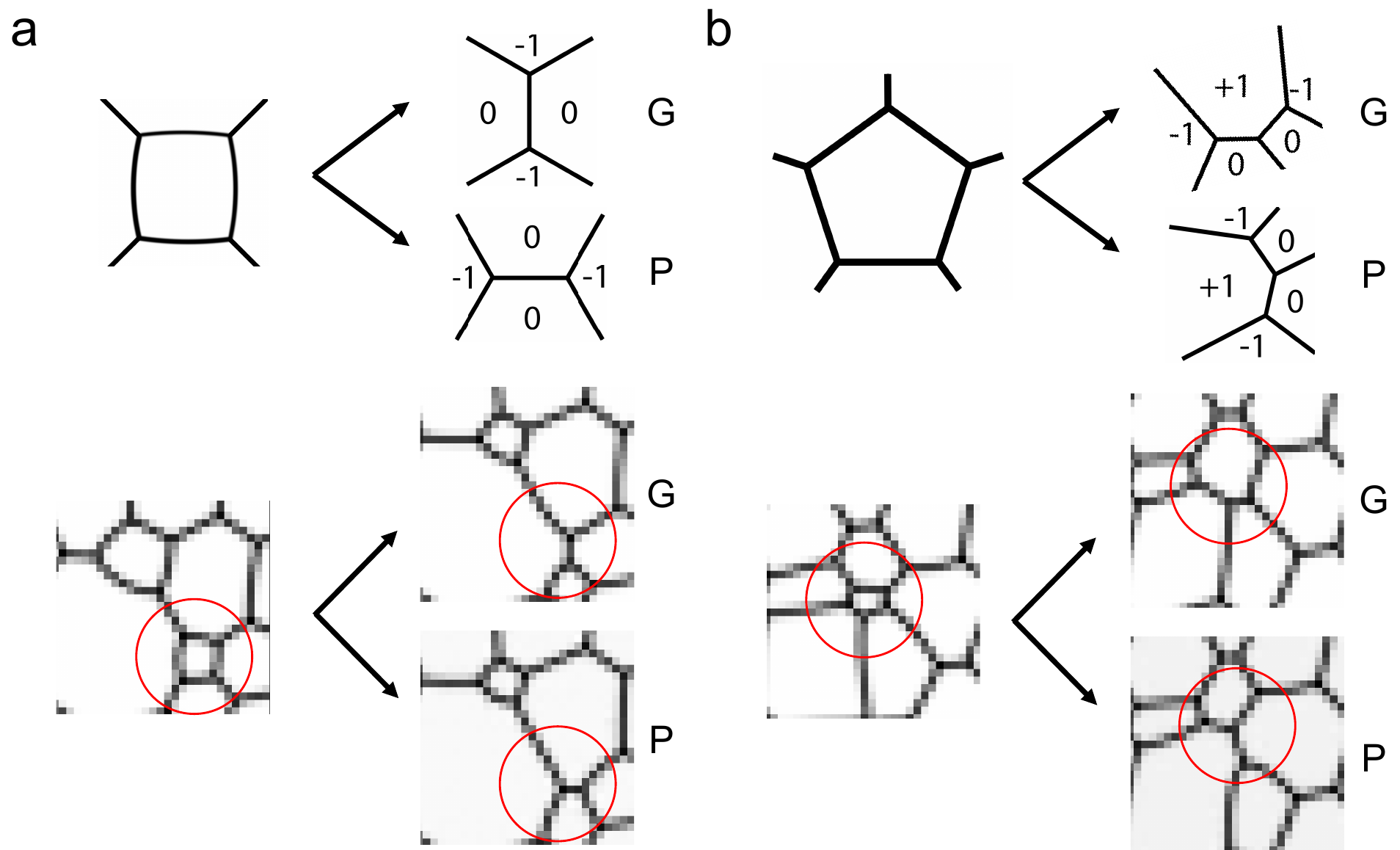}
	\caption{ \textbf{Examples of the bifurcation of grain boundary connectivity upon grain disappearance.} 
		Due to bifurcation, RNN predictions (P) differ from the ground truth (G) after the disappearance of a four-sided grain in \textbf{a.} or a five-sided grain in \textbf{b.} Top: schematics of the local topological changes that occur in predictions vs ground truth. Numbers inside each grain indicate the change to the number of sides after grain disappearance. Bottom: corresponding images from ground truth and RNN output. Red circles highlight the regions where the grain boundary connectivity bifurcates.  
		}
\end{figure}

\begin{figure}[!thp]
	\centering
	\includegraphics[width=0.65\columnwidth]{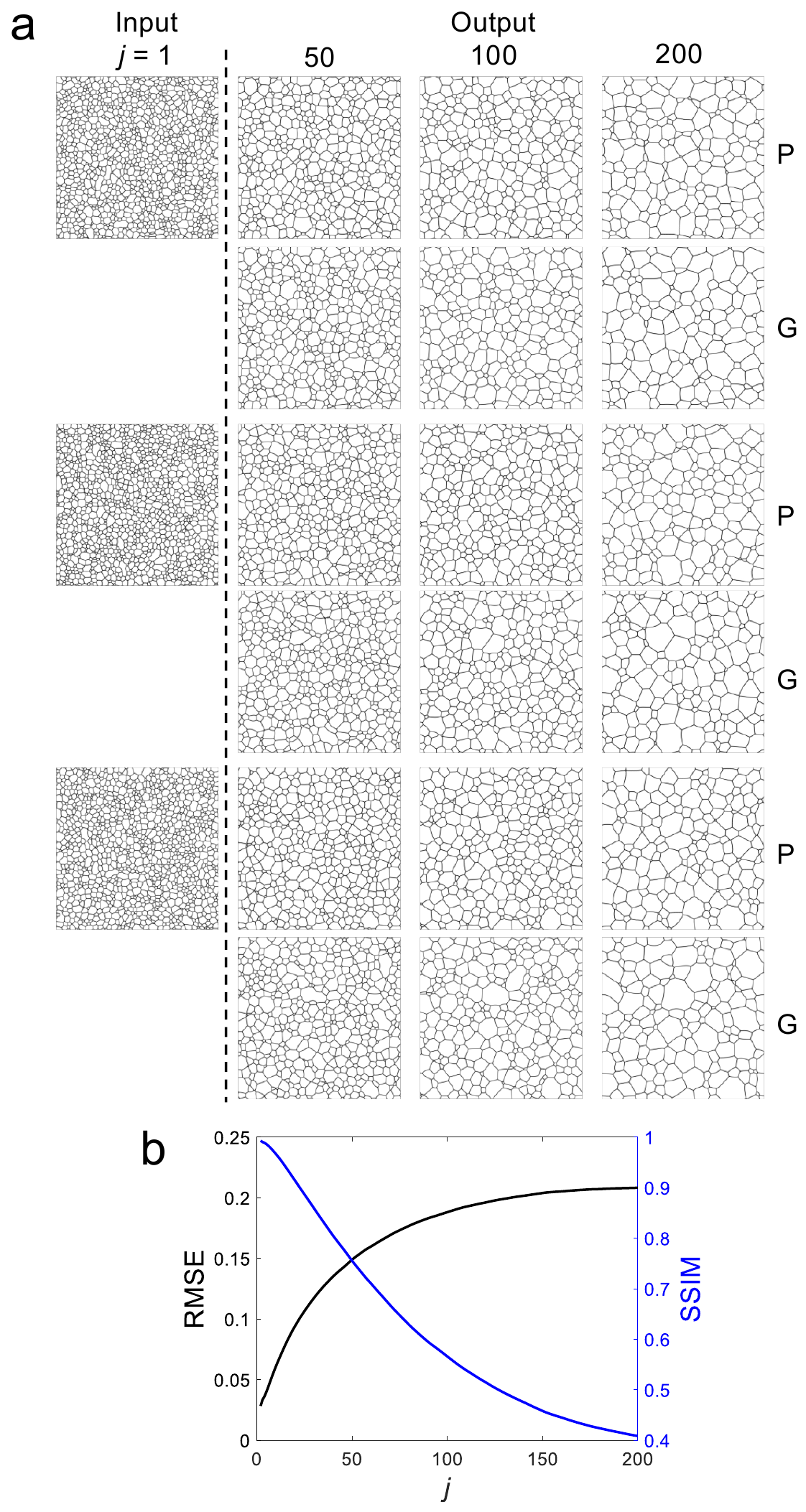}
	\caption{ \textbf{Prediction of grain growth in a larger 256$\times$256-pixel system using the RNN model trained on 64$\times$64-pixel image data.}
		\textbf{a.} Examples of RNN predictions (P) vs ground truth (G) from three testing cases, in which RNN outputs 200 frames based on 1 input frame. 
		\textbf{b.} RMSE (black) and SSIM (blue) of the predictions averaged over 50 testing cases.   
		}
\end{figure}

\begin{figure}[!thp]
	\centering
	\includegraphics[width=0.7\columnwidth]{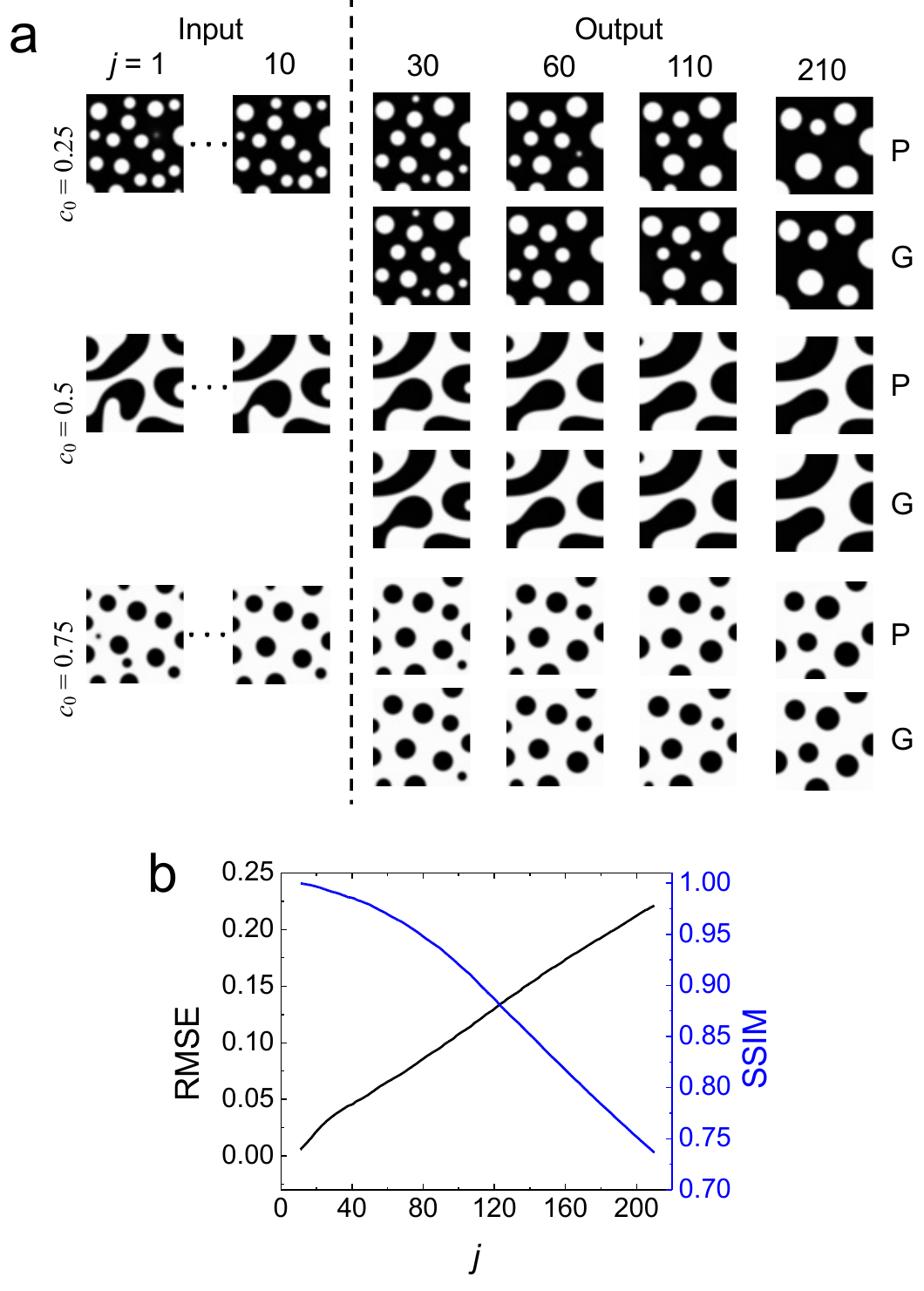}
	\caption{ \textbf{RNN predictions of spinodal decomposition based on 10 input frames.} 
		\textbf{a.} Examples of RNN predictions (P) vs ground truth (G) from three testing cases, in which RNN outputs 200 64$\times$64-pixel image frames based on 10 input frames, which are taken from the 41st to 50th frames of a simulation starting from a uniform mixture.
		\textbf{b.} RMSE (black) and SSIM (blue) of the predictions averaged over 510 testing cases.  
		}
\end{figure}

\begin{figure}[!thp]
	\centering
	\includegraphics[width=1\columnwidth]{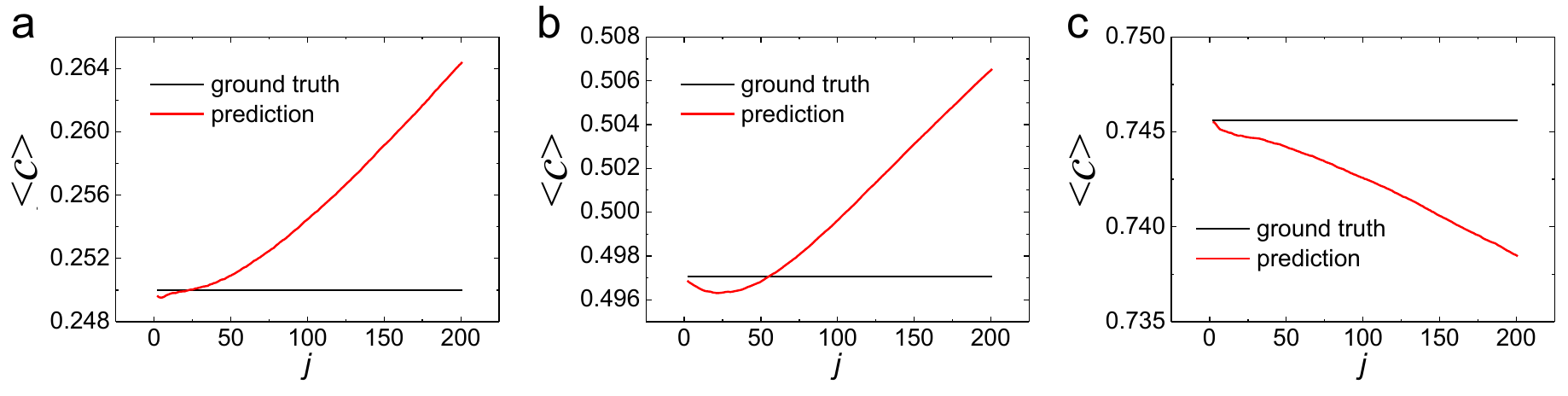}
	\caption{ \textbf{Evolution of the average system concentration $\langle c\rangle$ in RNN output frames}. 
		$\langle c\rangle$ is averaged over 170 testing cases with $c_0$ = 0.25 in \textbf{a}, 0.5 in \textbf{b} and 0.75 in \textbf{c}, respectively.  
		}
\end{figure}

\begin{figure}[!thp]
	\centering
	\includegraphics[width=0.9\columnwidth]{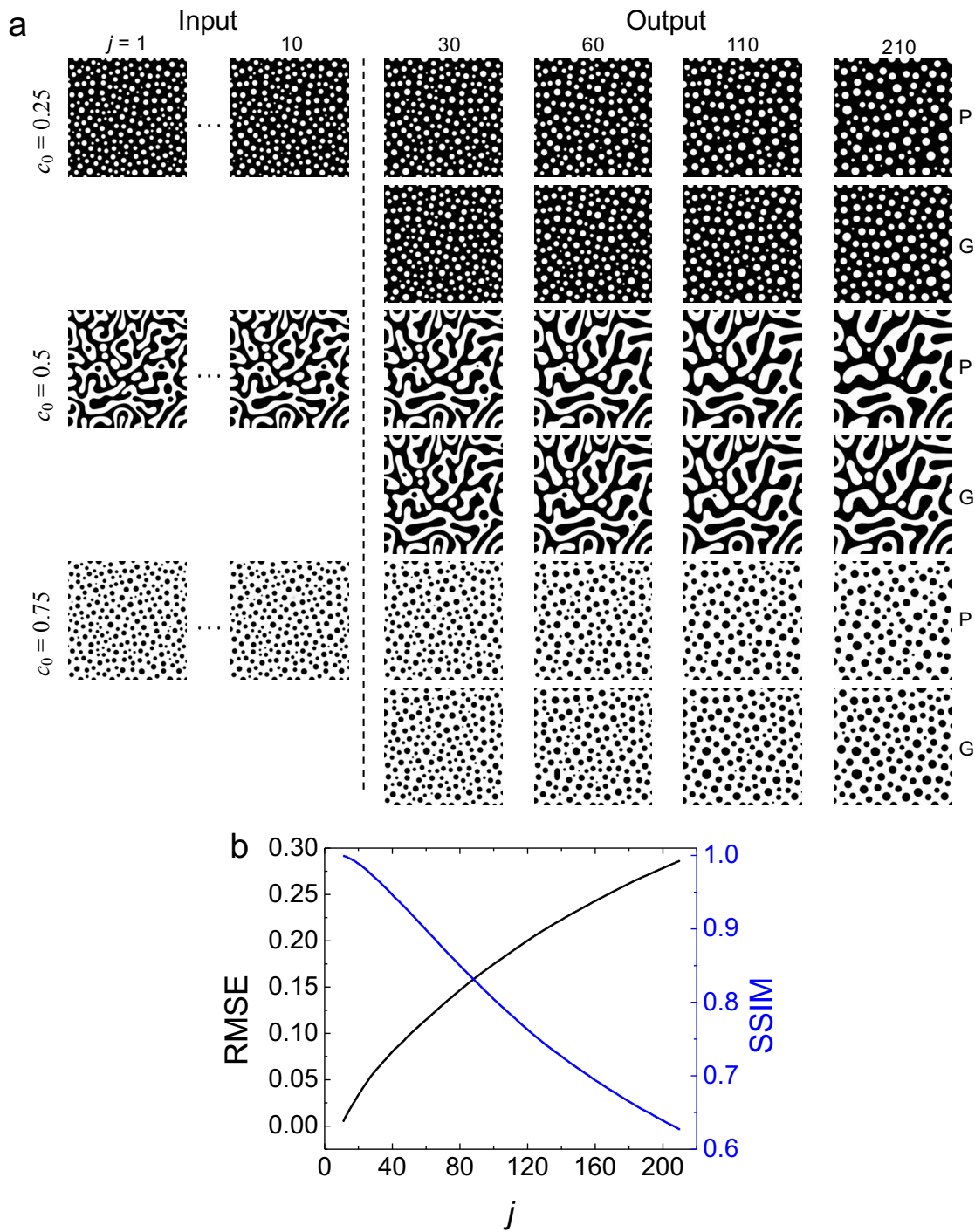}
	\caption{ \textbf{Prediction of spinodal decomposition in a larger 256$\times$256-pixel system using the RNN model trained on 64$\times$64-pixel image data.}
		\textbf{a.} Examples of RNN predictions (P) vs ground truth (G) from three testing cases, in which RNN outputs 200 256$\times$256-pixel frames based on 10 input frames. 
		\textbf{b.} RMSE (black) and SSIM (blue) of predictions averaged over 50 testing cases.  
		}
\end{figure}

\begin{figure}[!thp]
	\centering
	\includegraphics[width=0.6\columnwidth]{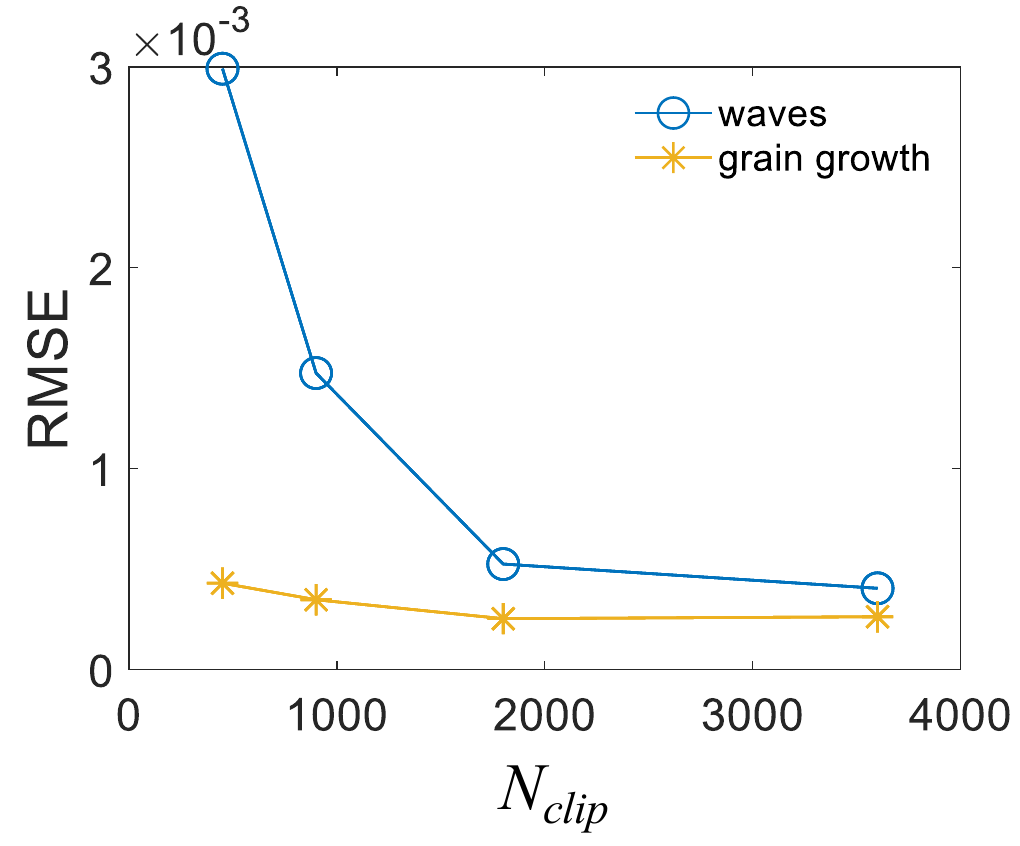}
	\caption{ \textbf{Dependence of the validation error on the number of training clips $N_{clip}$ in the application of RNN to predicting plane wave propagation and grain growth.}
		}
\end{figure}